\documentclass[draftcls, 12pt,onecolumn,oneside,compress]{IEEEtran}
\usepackage{times,amsmath,color,amssymb,graphicx,epsfig,cite,psfrag,subfigure,algorithm,balance}
\usepackage{amsfonts,pifont,enumerate,cases}
\usepackage{mathrsfs} 
\usepackage[table]{xcolor} 
\usepackage{verbatim} 
\usepackage{bm}
\usepackage{geometry}
\DeclareMathOperator*{\argmax}{argmax}
\DeclareMathOperator*{\argmin}{argmin}

\newcommand{\sinc}{\text{sinc}}

\begin{document}
\title{Channel Estimation for TDD/FDD Massive MIMO Systems with Channel Covariance Computing}
\author{Hongxiang Xie, Feifei Gao,  Shi Jin, Jun Fang, and Ying-Chang Liang
\thanks{H. Xie and F. Gao are with Tsinghua National Laboratory for Information Science and Technology (TNList) Beijing 100084, P. R. China (email: xiehx14@mails.tsinghua.edu.cn, feifeigao@ieee.org). S. Jin is with the National Communications Research Laboratory, Southeast University, Nanjing 210096, P. R. China (email: jinshi@seu.edu.cn). J. Fang is with the National Key Laboratory of Science and Technology on Communications, University of Electronic Science and Technology of China, Chengdu 611731, P. R. China (email: JunFang@uestc.edu.cn).
Y.-C. Liang is with the University of Sydney, NSW 2006, Australia, and also with the University of Electronic Science and Technology of China, Chengdu 611731, P.R. China (email: liangyc@ieee.org). }
}
\maketitle
\thispagestyle{empty}
\vspace{-15mm}
\begin{abstract}
In this paper, we propose a new channel estimation scheme for TDD/FDD massive MIMO systems by reconstructing\footnote{Sometimes called as covariance prediction
or covariance fitting.} uplink/downlink channel covariance matrices (CCMs) with the aid of array signal processing techniques. Specifically, the angle information and power angular
spectrum (PAS) of each multi-path  channel is extracted from the instantaneous uplink channel state information (CSI). Then, the uplink CCM is reconstructed and can be used to improve the uplink channel estimation
without any additional training cost.  In virtue of angle reciprocity as well as PAS reciprocity between uplink and downlink channels, the downlink CCM could also be inferred with a similar approach even for FDD massive MIMO systems. Then, the downlink instantaneous CSI can be obtained by training towards the dominant eigen-directions of each user.  The proposed strategy is applicable for any kind of PAS distributions and array geometries.
Numerical results  are provided to demonstrate the superiority of the proposed methods over the existing ones.
\end{abstract}

\vspace{-5mm}

\section{Introduction}\label{sec:intro}

Large-scale multiple-input multiple-output (MIMO) or ``massive MIMO'' is
a new technique that employs hundreds or even thousands of antennas at base station (BS)
to simultaneously serve multiple users and has been widely investigated for its numerous merits, such as
high spectrum and energy efficiency, high spatial resolution, and simple
transceiver design \cite{marzetta,scalingupMIMO, MIMOnextgen}.

To embrace all these potential gains, the accurate channel state
information (CSI) between BS and users is a prerequisite.
The CSI acquisition has been recognized as a very challenging task
for massive MIMO systems, due to the high dimensionality of channel matrices
as well as the resultant uplink pilot contamination, overwhelming downlink
training overhead, prohibitive computational complexity and so on \cite{marzetta}.
To overcome these challenges, many research works \cite{Caire,yin,BDMA} on massive MIMO channel estimation
have resorted to the channel statistics, e.g., channel correlation or channel
covariance matrix (CCM), and are built on  the facts that the finite scattering propagation
environments \cite{heath_mmwavereview} as well as the high correlation between compact antenna elements make high-dimensional CCMs sparse or low-rank. That is, CCMs could be utilized to project the high-dimensional channels onto
low-dimensional subspaces and thus reduce the effective channel dimensions.
Compared to the compressive sensing approaches which try to
recover CSI from fewer sub-Nyquist sampling points \cite{DaiTSP,CSIT} or the angle-space
  methods which exploit the intrinsic angle parameters \cite{JSAC,tvt,access},  CCM based channel
estimation
would offer additional statistical knowledge about the channel variants
and therefore achieve much better channel estimation accuracy.
Furthermore, space-time preprocessing at BS, like (hybrid) transmitter beamforming \cite{heath_hybridprecoding,caire_SubspaceEstimation},
channel prediction and  signal-to-noise ratio (SNR) prediction,
can benefit from the knowledge of CCMs as well.

It is worthy mentioning that the CCM  is different from signal covariance
where the latter can be easily obtained from the accumulation of the received signals
whereas the former can only be acquired from the accumulation of channel estimates. For massive MIMO system, unfortunately,
the number of channel estimates to construct CCMs increases linearly with the array size,
making the accuracy or even the feasibility of the CCM acquisition questionable.
Especially, it is much more difficult to obtain the downlink CCMs for all users as the cost
of training and feedback is hardly affordable.
To release the heavy burden of downlink CCM acquisition,
the common practice is to utilize the channel reciprocity
in time division duplexing (TDD) systems,  where the downlink CCM can be
immediately obtained from its uplink counterpart. However, channel reciprocity is not applicable for frequency division duplexing (FDD) systems.
Some works have proposed downlink CCM reconstruction (or inference) for conventional FDD MIMO systems by
exploring the structure of antenna array.
For example, the authors of \cite{Hugl&Bonek} extracted the power angular spectrum (PAS)
from measured uplink CCMs and then transformed it into the estimated downlink correlation, assuming
PAS reciprocity  between uplink and downlink channels.
In \cite{Liang}, the uplink CCM was directly used to predict its
downlink counterpart through Fourier transform without explicit calculation
of the PAS, but still relying on the assumption that uplink and downlink PASs are identical.
Based on the observed uplink covariance, authors of \cite{CCMinterpolation} proposed an inference process
to interpolate the FDD downlink CCMs over a Riemannian space after getting a dictionary of
uplink/downlink CCMs pairs at certain frequency points through
preamble training.
However, all these methods  \cite{Hugl&Bonek,Liang,CCMinterpolation} require the knowledge of uplink CCMs without considering
their acquisition difficulties.  Moreover, directly applying \cite{Hugl&Bonek,Liang,CCMinterpolation} to   massive MIMO scenarios may not
be a good choice, since they do not exploit
the special characteristics of ``massive'' number of antennas.  Being aware of the high angular resolution
of large-scale antenna arrays, the authors of \cite{Fang}  formulate CCM as an integration of certain functions over the
angular spread (AS) of signals. Nevertheless, the result in \cite{Fang}
is only valid for the  uniform distribution of PAS and has to assume  very  narrow AS as well as the  uniform linear array (ULA), which is far from a general discussion.

In this paper,  we propose a new channel estimation scheme for TDD/FDD massive MIMO
systems with inferred uplink/downlink CCMs from array signal processing techniques.
Specifically, the angle and PAS parameters of each multi-path channel
are first extracted from one instantaneous uplink channel estimate.
Then uplink CCMs are then reconstructed with these angle and PAS parameters based on the structures of large-scale antenna arrays.
With the reconstructed uplink CCMs, the uplink channel can be re-estimated in a better shape without any additional training cost.
Thanks to the angle reciprocity as well as PAS similarity between uplink/downlink channels,
the downlink CCMs can be  inferred even for FDD massive MIMO systems,
and then are utilized to enable the eigen-beamforming for downlink CSI estimation.
Comparing with existing CCM reconstruction methods \cite{Hugl&Bonek,Liang,CCMinterpolation,Fang,OFDM_wiener,sigcom},
the proposed offers following several difference and benefits:
\begin{enumerate}
\item The proposed method fully exploits the high spatial resolution of massive MIMO.
  \item The proposed method reconstructs CCM with only one instantaneous uplink channel estimate, from which
  channel estimation performance  is self-enhanced without any additional training overhead.
  \item The proposed method is suitable for both TDD and FDD massive MIMO systems.
  \item The proposed method is applicable for any kinds of PAS distributions and array geometries.
\end{enumerate}

The rest of the paper is organized as follows. In section II,
the system model as well as the representations and relationships of uplink/downlink CCMs are described.
The proposed CCM reconstruction and inference scheme for TDD/FDD massive MIMO systems are presented in
section III, followed by simulations in section IV. Finally, conclusions are drawn in section V.

\textbf{Notations:} Vectors and matrices are denoted by boldface small and
capital letters;  the transpose, Hermitian,
inverse, and pseudo-inverse of the matrix ${\mathbf A}$ are denoted
by ${\mathbf A}^T$, ${\mathbf A}^H$, ${\mathbf
A}^{-1}$ and ${\mathbf A}^\dag$, respectively;
$[{\mathbf A}]_{i,j}$ is the $(i,j)$-th entry of
${\mathbf A}$ and the entry index of vector and matrix starts from $0$;
${\mathbf I_M}$ is an $M\times M$
identity matrix; $\mathbb{E}\{\cdot\}$ is the statistical
expectation; $\triangleq$ denotes new definition;
$\left[\mathbf h\right]_{\mathcal{Q},:}$ indicates
the sub-vector of $\mathbf h$ by keeping the elements indexed by $\mathcal{Q}$;
$\left[\mathbf H\right]_{:,\mathcal{Q}}$ denotes the
sub-matrix of $\mathbf H$ by collecting the columns indexed by $\mathcal{Q}$,
and $\left[\mathbf H\right]_{\mathcal{Q},:}$ denotes the
sub-matrix of $\mathbf H$ by collecting the rows indexed by $\mathcal{Q}$.

\section{Problem Formulation}

\begin{figure}[t]
      \centering
     \includegraphics[width=100mm]{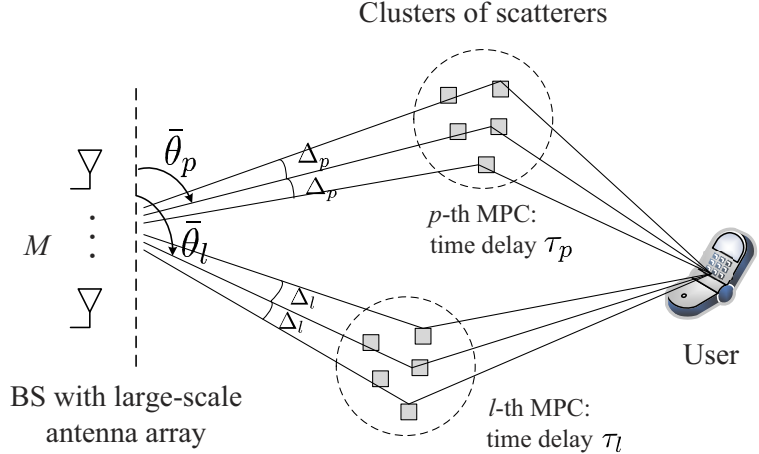}
     \caption{ The clustered response model for multi-path environment, where each multi-path component
   consists of a continuum of indistinguishable rays. }
     \label{fig:systemmodel}
\end{figure}
\subsection{System Model}

Let us consider a general finite scattering environment for massive MIMO systems shown in Fig. \ref{fig:systemmodel},
where the BS is equipped with an $M\gg 1$ antennas and is  located at an elevated position with few surrounding scatterers. The
channel between BS and a user experiences frequency selective fading and
the multi-path environment is characterized by the clustered response model \cite{Tse_fundamentalWireless}. The scatterers and reflectors are not located at all directions from the BS or the users but are grouped into several
clusters with each cluster bouncing  off a multi-path component (MPC) comprised of a continuum of simultaneous rays.

Without loss of generality, we suppose
the propagation from a destined user to BS will go through $P$ resolvable independent multi-paths,
and the mean AOA as well as the AS of
the $p$-th MPC are denoted by $\bar{\vartheta}_{p}$ and $\Delta_{p}$, respectively.
Then the corresponding $M \times 1$ uplink channel for the $p$-th MPC can be expressed by integrating over the
incident angular region as
\begin{equation}\label{equ:channelmodel}
\mathbf h_{p}=\int_{\vartheta\in\mathcal{A}_{p}}\alpha(\vartheta)\mathbf a(\vartheta)d\vartheta
=\int_{\bar{\vartheta}_p-\Delta_p}^{\bar{\vartheta}_p+\Delta_p}|\alpha(\vartheta)|e^{j\phi(\vartheta)}\mathbf a(\vartheta)d\vartheta,
\end{equation}
where $\mathcal{A}_{p}\triangleq[\bar{\vartheta}_{p}-\Delta_{p}, \bar{\vartheta}_{p}+\Delta_{p}]$
and $\alpha(\vartheta)\triangleq |\alpha(\vartheta)|e^{j\phi(\vartheta)}$ with $|\alpha(\vartheta)|$ and $\phi(\vartheta)$ representing the attenuation (amplitude) and phase of the incident signal ray coming from AOA $\vartheta$, respectively.
Moreover, $\mathbf a(\vartheta)$ is the array manifold vector (AMV) or array response vector, whose expression is dependent on
the specific array geometries. When a ULA is deployed, there is
\begin{equation}\label{equ:AMV}
\mathbf a(\vartheta)=\frac{1}{\sqrt{M}}\left[1,e^{-j\chi\cos(\vartheta)}, e^{-j2\chi\cos(\vartheta)},\ldots,e^{-j(M-1)\chi\cos(\vartheta)}\right]^T,
\end{equation}
where $\chi\triangleq  2\pi f d/c$, $d$ denotes the antenna spacing, $f$ is the carrier frequency and $c$ is the speed of light.
Obviously, $\mathbf a(\vartheta)$ is dependent on the signal carrier frequency $f$.
Then the channel frequency response of the user can be written as \cite{Li_ofdm}
\begin{align}\label{equ:MPChannel}
  \mathbf h(f)&=\sum_{p=1}^P \mathbf h_{p}\cdot e^{-j2\pi f\tau_{p}} =\sum_{p=1}^P
  e^{-j2\pi f\tau_{p}}\int_{\mathcal{A}_{p}}\alpha(\vartheta)\mathbf a(\vartheta)d\vartheta ,
\end{align}
where $\tau_{p}$ is the time delay of the $p$-th MPC.

From above channel model, the $M\times M$ CCM of the $p$-th MPC $\mathbf h_p$ can be written as
\begin{align}\label{equ:R}
  \mathbf R_p=\int_{\vartheta\in\mathcal{A}_p}\mathbb{E}\{|\alpha(\vartheta)|^2\}\mathbf a(\vartheta)\mathbf a^H(\vartheta) d\vartheta
  =\int_{\bar{\vartheta}_p-\Delta_p}^{\bar{\vartheta}_p+\Delta_p} S_p(\vartheta)\mathbf a(\vartheta)\mathbf a^H(\vartheta) d\vartheta,
\end{align}
where $S_p(\vartheta)\triangleq \mathbb{E}\{|\alpha(\vartheta)|^2\}$ denotes the PAS of $\mathbf h_p$, characterizing the channel power distribution
of the $p$-th MPC in angular domain.  Obvious, $\mathbf R_p$ is determined by the values of central AOA $\bar{\vartheta}_p$, AS $\Delta_p$, PAS $S_p(\vartheta)$, as well as the AMV $\mathbf a(\vartheta)$.
Accordingly, if ULA is deployed, the $(m,n)$-th element of $\mathbf R_p$ is denoted by
\begin{align}\label{equ:Rmn}
  [\mathbf R_p]_{m,n}=\frac{1}{M}\int_{\bar{\vartheta}_p-\Delta_p}^{\bar{\vartheta}_p+\Delta_p} S_p(\vartheta) e^{-j(m-n)\chi\cos(\vartheta)} d\vartheta.
\end{align}


\subsection{Representation and Expansion of CCM}

Since $S_p(\vartheta)$ is also a function of the angle parameters $\bar{\vartheta}_p$ and $\Delta_p$,
the estimation of $\mathbf R_p$ actually boils down to two sub-problems: angle estimation and PAS estimation.
For angle acquisition, many canonical means such as various MUSIC \cite{music} or ESPRIT methods \cite{esprit},
or some emerging approaches for large-scale antenna arrays like
compressive sensing (CS) \cite{superresolution1,superresolution2} or discrete Fourier transform (DFT) \cite{tvt} are available, whose detailed exposition will be left in  Section III.
Let us here  focus on the analysis of PAS.

\subsubsection{Known PAS Distribution}
If the prior distribution of $S_p(\vartheta)$ is available, then the expression of $[\mathbf R_p]_{m,n}$ in
\eqref{equ:Rmn} could be possibly simplified to a function of $\bar{\vartheta}_p$ and $\Delta_p$.
For example, if $S_p(\vartheta)$ is uniformly distributed with mean AOA $\bar{\vartheta}_p$ and AS $\Delta_p$, i.e.,  $S_p(\vartheta)=\frac{1}{2\Delta_p} $
for $\forall\vartheta \in[\bar{\vartheta}_p-\Delta_p,\bar{\vartheta}_p+\Delta_p]$ \cite{Fang}, then
the $(m,n)$-th entry of $\mathbf R_p$ can be expressed as
\begin{subequations}\label{equ:uniformP}
\begin{align}
    [\mathbf R_p]_{m,n}&=\frac{1}{2M\Delta_p}\int_{\bar{\vartheta}_p-\Delta_p}^{\bar{\vartheta}_p+\Delta_p} e^{-j(m-n)\chi\cos(\vartheta)} d\vartheta \label{equ:6a} \\
    &\approx\frac{1}{2M\Delta_p}e^{-j(m-n)\chi\cos(\bar{\vartheta}_p) }\int_{ -\Delta_p}^{ \Delta_p} e^{j(m-n)\chi\sin(\bar{\vartheta}_p)\vartheta}d\vartheta \notag\\
    &=\frac{1}{M}e^{-j\chi_{mn}\cos(\bar{\vartheta}_p) }\sinc\left(\chi_{mn} \Delta_p \sin(\bar{\vartheta}_p)\right),\label{equ:6b}
\end{align}
\end{subequations}
where $\chi_{mn}\triangleq(m-n)\chi$ and $\sinc(x)\triangleq\sin x/ x$ denotes the sinc function. The second approximation equation
derives from the Taylor expansion of $\cos(\vartheta)$ at $\bar\vartheta$, namely, $\cos(\vartheta)\approx \cos(\bar\vartheta)-(\vartheta-\bar\vartheta)\sin\bar\vartheta$.
Similarly, if $S_p(\vartheta)$ is modeled by Laplacian distribution \cite{Pedersen_PAS} as
\begin{equation}\label{equ:Laplacianpdf}
S_p(\vartheta)=\frac{1}{\sqrt{2}\Delta_p}e^{-\frac{\sqrt{2}|\vartheta-\bar{\vartheta}_p|}{\Delta_p}},
\end{equation}
then there is
\begin{subequations}\label{equ:LaplacianP}
\begin{align}
    [\mathbf R_p]_{m,n}&=\int_{\bar{\vartheta}_p-\Delta_p}^{\bar{\vartheta}_p+\Delta_p} \frac{1}{\sqrt{2}M\Delta_p}e^{-\frac{\sqrt{2}}{\Delta_p}|\vartheta-\bar{\vartheta}_p|-j\chi_{mn}\cos(\vartheta)} d\vartheta  \label{equ:8a}\\
    &\approx\frac{1}{\sqrt{2}M\Delta_p}e^{-j\chi_{mn}\cos(\bar{\vartheta}_p) }\int_{ -\Delta_p}^{ \Delta_p} e^{-\frac{\sqrt{2}|\vartheta|}{\Delta_p}+j\chi_{mn}\sin(\bar{\vartheta}_p)\vartheta} d\vartheta \notag\\
    &=\frac{1}{\sqrt{2}M}e^{-j\chi_{mn}\cos( \bar{\vartheta}_p)}\frac{1}{2+\chi_{mn}^2\Delta_p^2\sin^2(\bar\vartheta)}\left\{2\sqrt{2}
    \left(1-e^{-\sqrt{2}}\cos\left(\chi_{mn}\Delta_p\sin(\bar\vartheta)\right)\right)\right. \notag\\
    &\left.+2e^{-\sqrt{2}}\chi_{mn}^2\Delta_p^2\sin^2(\bar\vartheta)~\sinc\left(\chi_{mn}\Delta_p\sin(\bar\vartheta)\right)\right\}.\label{equ:8b}
\end{align}
\end{subequations}

Note that the closed-form equations \eqref{equ:6b} and \eqref{equ:8b} are
the approximation of \eqref{equ:6a} and \eqref{equ:8a}, respectively,
under the condition of very narrow AS.  That is to say, if the AS is not small enough, we
should refer back to \eqref{equ:6a} and \eqref{equ:8a} for the true CCMs reconstruction. In this case,
the basic Monte Carlo method may be adopted for the integral process and  thus we refer to this procedure as the \emph{Monte Carlo integral CCM
reconstruction (MC-iCCM)  }.  In comparison, the closed-form equations \eqref{equ:6b} and \eqref{equ:8b} are
referred to as the \emph{Closed-form integral CCM reconstruction  (CF-iCCM)  }.

\subsubsection{Unknown  PAS Distribution}
Let us then consider a more practical  case where the prior knowledge of the distribution of $S_p(\vartheta)$ is unavailable. Under this circumstance, $S_p(\vartheta)$ could not be explicitly expressed by a given probability density function and
thereby $[\mathbf R_p]_{m,n}$ may not be simplified like equations \eqref{equ:uniformP} or \eqref{equ:LaplacianP}.

Without any prior knowledge of $S_p(\vartheta)$,  one should first estimate $S_p(\vartheta)$.  With limited number of observations over antennas, we need first  approximate $\mathbf R_p$
by an appropriate basis expansion model (BEM), or to approximate the integral
of \eqref{equ:R} by the sum of discrete points extracted from the angular integral region.
For example, the whole spatial space is often discretized evenly into $M$  blocks,
each with the width of $\frac{2\pi}{M}$, and
then the basis vectors could be chosen as the columns of an $M\times M$ DFT matrix when ULA is deployed at BS.
In this case,  the continuous PAS function in \eqref{equ:R} should be approximated  by $M$ discrete expansion coefficients,
denoted as $\{S_{p,l}\}$ for $l=0,1,\ldots, M-1$. Then, there is
\begin{align}\label{equ:Mpk}
  \mathbf R_p=\int_{\bar{\vartheta}_p-\Delta_p}^{\bar{\vartheta}_p+\Delta_p} S_p(\vartheta)\mathbf a(\vartheta)\mathbf a^H(\vartheta) d\vartheta
\approx\sum_{l=0}^{M-1} S_{p,l}\mathbf f_l\mathbf f_l^H=\mathbf F\text{diag}\left\{S_{p,0},S_{p,1},\ldots,S_{p,M-1}\right\}\mathbf F^H,
\end{align}
where $\mathbf F$ is the $M\times M$ normalized DFT matrix whose $(p,q)$-th element is  $\left[\mathbf F\right]_{p,q}=e^{-j\frac{2\pi}{M}pq}/\sqrt{M}$
and $\mathbf f_l$ is the $l$-th column of $\mathbf F$. Therefore, $S_{p,l}$ is given as
\begin{align}\label{equ:Spl}
  S_{p,l}&\approx\mathbf f_l^H\mathbf R_p\mathbf f_l
  =\int_{\bar{\vartheta}_p-\Delta_p}^{\bar{\vartheta}_p+\Delta_p} S_p(\vartheta)\mathbf f_l^H\mathbf a(\vartheta)\mathbf a^H(\vartheta)\mathbf f_l d\vartheta.
\end{align}
The estimation of PAS is  thus equivalent to estimating the $M$ expansion coefficients $\{S_{p,l}\}_{l=1}^M$.

A more accurate way other than using  DFT expansion is to approximate $\mathbf R_p$ with non-orthogonal basis vectors within the actual incident AS. For instance, without any prior knowledge of $S_p(\vartheta)$, a natural idea is to sample $S_p(\vartheta)$ at discrete points, e.g., $\{S_p(\vartheta_l)| \vartheta_l\in \mathcal{A}_p, l=0,1,\ldots,L-1\}$. Then the approximation of $\mathbf R_p$ could be expressed as
\begin{align}\label{equ:discreteR}
  \mathbf R_p=\int_{\bar{\vartheta}_p-\Delta_p}^{\bar{\vartheta}_p+\Delta_p}S_p(\vartheta)\mathbf a(\vartheta)\mathbf a^H(\vartheta)d\vartheta
  \approx \sum_{l=0, \forall\vartheta_l\in\mathcal{A}_p}^{L-1} S_p(\vartheta_l)\mathbf a(\vartheta_l)\mathbf a^H(\vartheta_l).
\end{align}
It is apparent that $S_p(\vartheta_l)$ is slightly different from $S_{p,l}$ in \eqref{equ:Spl}.
As we increase the number of sum terms $L$ in \eqref{equ:discreteR}, $S_p(\vartheta_l)$'s is a
more accurate approximation of the continuous PAS $S_p(\vartheta)$ and thus renders a more
accurate CCM approximation.

In these cases, to reconstruct CCMs, not only the central AOA $\bar{\vartheta}_p$ and AS $\Delta_p$ but also the discrete samples of PAS  $S_p(\vartheta_l)$'s are needed.

\subsection{Inferring Downlink CCMs from  Uplink CCMs}
Since the AMV $\mathbf a(\vartheta)$ is dependent on the signal carrier frequencies, downlink CCMs will differ from their uplink counterparts in FDD systems.
Denote the uplink and downlink signal carrier frequencies by $f_u $ and $f_d$ respectively.
Meanwhile, let us rewrite the uplink AMV $\mathbf a(\vartheta)$ and CCM  $\mathbf R_p$ as $\mathbf a_u(\vartheta)$ and $\mathbf R_p^u$, respectively
and represent their downlink ones accordingly as $\mathbf a_d(\vartheta)$ and $\mathbf R_p^d$. Then, there are
\begin{align}
  \mathbf R_p^u&=\int_{\vartheta\in\mathcal{A}^u_p} S^u_p(\vartheta)\mathbf a_u(\vartheta)\mathbf a_u^H(\vartheta) d\vartheta,\\
  \mathbf R_p^d&=\int_{\vartheta\in \mathcal{A}^d_p} S^d_p(\vartheta)\mathbf a_d(\vartheta)\mathbf a_d^H(\vartheta) d\vartheta,
\end{align}
where $\mathcal{A}_p^d$ is the downlink angle of departure (AOD) interval at BS and $S_p^d(\vartheta)$ is the downlink PAS.
Considering the structure of  $\mathbf a_u(\vartheta)$  and $\mathbf a_d(\vartheta)$, the following transformation holds:
\begin{align} \label{equ:uldlAMV}
  \mathbf a_d(\vartheta) = \mathbf \Theta(\vartheta)\mathbf a_u(\vartheta),
\end{align}
with $\mathbf \Theta(\vartheta)$ defined as
\begin{align}
  \mathbf \Theta(\vartheta)\triangleq\text{diag}\left\{1,e^{-j\frac{2\pi d}{c}(f_d-f_u)\cos(\vartheta)},e^{-j\frac{2\pi d}{c}(f_d-f_u)2\cos(\vartheta)},\ldots, e^{-j\frac{2\pi d}{c}(f_d-f_u)(M-1)\cos(\vartheta)}\right\}.
\end{align}

To proceed, we adopt the following assumptions from  many theoretical works \cite{Hugl&Bonek,Liang,sigcom,HuiLiu,heath_mmwavereview} and measurement tests \cite{Bonek_RealTimeDoa,Imtiaz_DirectRecipr,Hugl_SpatialReciprocity,Ericsson,Liang_[GF]}:
\begin{enumerate}
\item The underlying angle parameters of uplink and downlink channels are reciprocal, i.e, $\mathcal{A}_p^d\approx\mathcal{A}^u_p$, as long as
the uplink and downlink frequency discrepancy is not big.
\item The PAS of uplink and downlink channels are reciprocal, except for a frequency dependent scalar $\mu$, i.e.,  $S_p^d(\vartheta)=\mu S^u_p(\vartheta)$. That is, the shapes of uplink/downlink PAS functions are the same.
\end{enumerate}

Based on these facts, the relationship between uplink/downlink CCMs can be derived  as
\begin{align}\label{equ:ULDLtrans}
  \mathbf R_p^d = \int_{\bar{\vartheta}_p-\Delta_p}^{\bar{\vartheta}_p+\Delta_p} \mu S^u_p(\vartheta)\mathbf \Theta(\vartheta)\mathbf a_u(\vartheta)\mathbf a_u^H(\vartheta)\mathbf\Theta^H(\vartheta) d\vartheta.
\end{align}
By combining \eqref{equ:discreteR} and \eqref{equ:uldlAMV}, the discrete approximation of \eqref{equ:ULDLtrans} is given by
\begin{align}\label{equ:ULDLtransdiscrete}
  \mathbf R_p^d \approx \sum_{l=0, \forall\vartheta_l\in\mathcal{A}^u_p}^{L-1} \mu S^u_p(\vartheta_l)\mathbf a_d(\vartheta_l)\mathbf a_d^H(\vartheta_l)
  =\sum_{l=0, \forall\vartheta_l\in\mathcal{A}^u_p}^{L-1}\mu S^u_p(\vartheta_l)\mathbf\Theta(\vartheta_l)\mathbf a_u(\vartheta_l)\mathbf a_u^H(\vartheta_l)\mathbf \Theta^H(\vartheta_l).
\end{align}
Following above analysis, it is clear that the uplink angle parameters and PAS could be used to infer the
downlink CCM, even in FDD systems.


\section{Proposed Channel Estimation Scheme}
\begin{figure}[t]
\centering
\includegraphics[width=150mm]{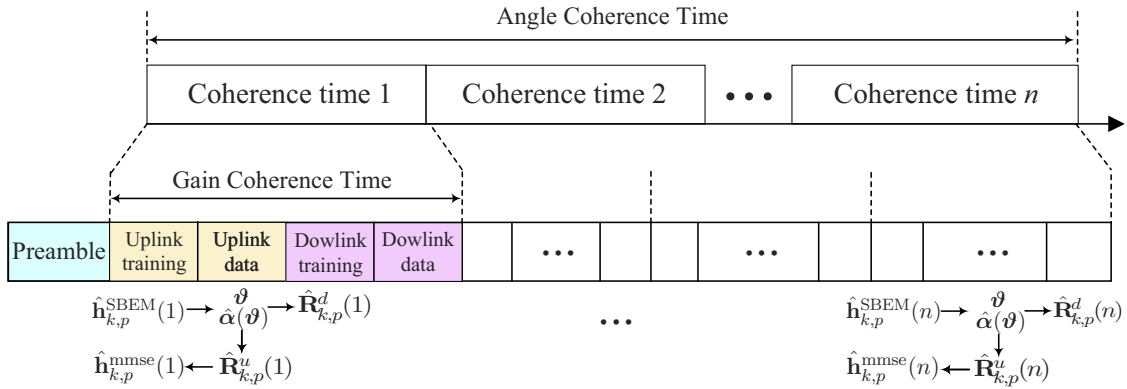}
\caption{ Transmission phase of our proposed scheme.
\label{fig:TransmissionPhase}}
\end{figure}

\begin{figure}[t]
\centering
\includegraphics[width=120mm]{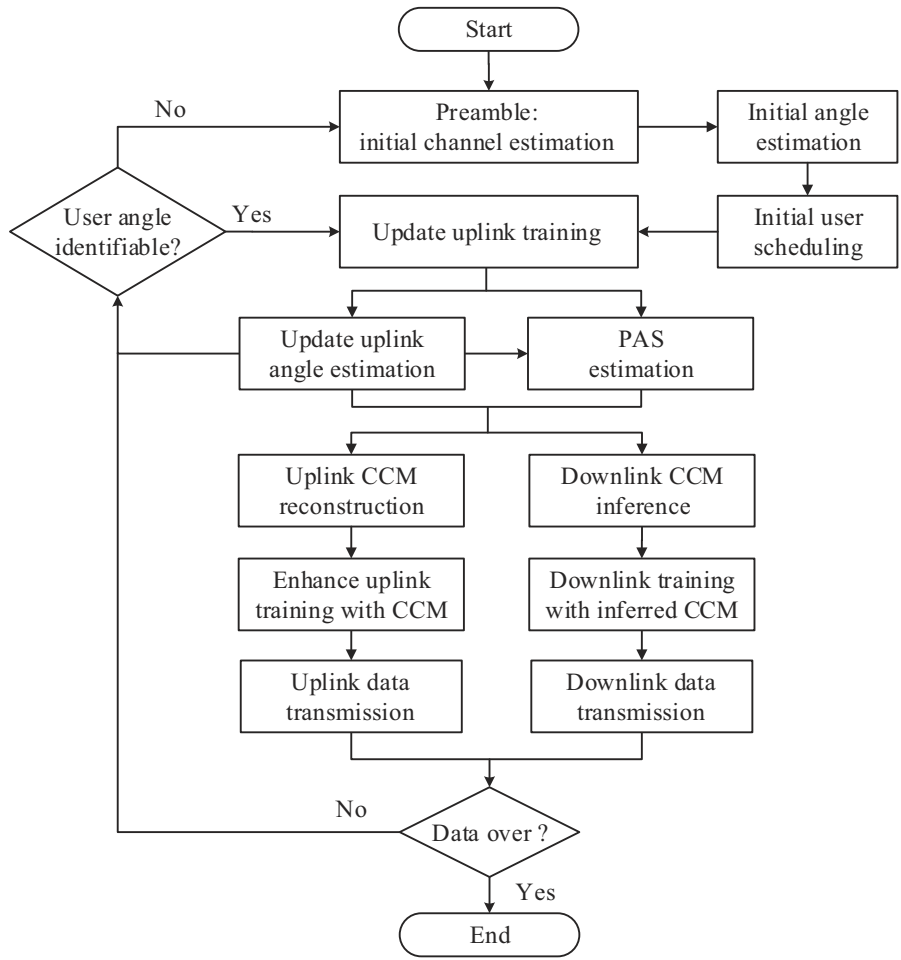}
\caption{ The flow chart of our proposed scheme.
\label{fig:cov_flowchart}}
\end{figure}

As shown in Fig. \ref{fig:TransmissionPhase},
the proposed transmission  starts from an uplink preamble
phase to obtain all users' initial AS information, which will be used for initial user grouping and scheduling.
In the subsequent transmissions, the instantaneous channel estimates are updated with limited pilots, and users' angle information are also dynamically updated. Then, the uplink/downlink CCMs are reconstructed and utilized to enhance the  channel estimation via minimum mean square error (MMSE) estimators.
The flow chart of the overall transmission scheme is presented in Fig. \ref{fig:cov_flowchart} for better understanding.

For ease of illustration,  we here will only focus on channel estimation for the $p$-th MPC of all users.
Meanwhile, we use an additional subscript $k$ to indicate all the symbols for user-$k$.

\subsection{Preamble For Initial Angle Estimation and User Scheduling}
During the preamble phase, all $K$ users adopt the conventional uplink training methods\footnote{The conventional training by assigning each user with one orthogonal training  may be costly but will be only sent once at the start of the transmission.}, e.g., least square (LS), to  estimate their initial CSI $\hat{\mathbf h}^{\text{ini}}_{k,p}$, $k=1,\ldots,K$. The next step is to extract the initial   $\mathcal{A}_{k,p}$ of each user from $\hat{\mathbf h}^{\text{ini}}_{k,p}$.

\subsubsection{Initial Angle Estimation}
Owing to the narrow AS, namely, $\Delta_{k,p}$ is small (especially, $\Delta_{k,p}$ is zero in mmWave scenarios \cite{heath_mmwavereview}), $\mathbf h_{k,p}$ in \eqref{equ:channelmodel} will exhibit sparsity or low-rank property in the angle domain and
thus it could be approximately represented by discretizing the integral of \eqref{equ:channelmodel} as
\begin{align}\label{equ:initialangleesti}
    \mathbf h_{k,p} = \mathbf A   \bm \alpha,
\end{align}
where $\mathbf A=[\mathbf a(\theta_0), \mathbf a(\theta_1),\ldots,\mathbf a(\theta_{L-1})]\in\mathbb{C}^{M\times L}$
with $\{\theta_l\}_{l=0}^{L-1}\in [-\pi,\pi]$ is an angle domain transform matrix.
Note that, here the possible range of unknown $\theta_l$ belongs to $[-\pi,\pi]$, different
from the range of true angles  $\vartheta\in\mathcal{A}_{k,p}$.
Moreover,  $\bm \alpha \in\mathbb{C}^{L\times 1}$ is the sparse representation to be determined and only a few components of $ \bm \alpha$
 corresponding to those angles inside $\mathcal{A}_{k,p}$ are nonzero.  Therefore, the nonzero components of $\bm \alpha$
 could facilitate the estimation of AOA interval $\mathcal{A}_{k,p}$. To this end, the initial angle estimation problem is formulated as
\begin{align}\label{equ:LS}
\left[ \hat{\bm \alpha},\{\hat{\theta}_l\}^{L-1}_{l=0}\right]=\argmin_{\bm \alpha,~ \{\theta_l\}^{L-1}_{l=0}}\ &\|\hat{\mathbf h}^{\text{ini}}_{k,p}-\mathbf A \bm \alpha\|^2.
\end{align}
To solve this problem, there are several common options. First, we can resort to maximum likelihood (ML) \cite{ML} method to search
for the optimal angle candidates $\{\hat{\theta}_l\}_{l=0}^{L-1}$. Second, if $L\gg M$, the super-resolution CS \cite{superresolution1,superresolution2} is  a popular
and suitable choice for such kind of sparse recovery problem. The spatial rotation enhanced DFT method from array signal
processing in \cite{tvt} is also a good choice.
Among the three options, ML and CS methods are capable of achieving high angle estimation accuracy.
Nevertheless, ML requires exhaustive search in the whole angle range $[-\pi,\pi]$, while CS relies on
non-linear optimization and iterations, both of which suffer from relatively high complexity. By contrast,
the spatial rotation enhanced DFT method achieves a nice tradeoff between angle estimation accuracy and computational
complexity. In this paper,  we adopt the third option for angle acquisition.

According to \cite{tvt}, $\mathbf A $ is chosen as $\mathbf A=\mathbf \Phi^H(\psi)\mathbf F^H$, with
the spatial rotation matrix $\mathbf \Phi(\psi)$ defined as
$\mathbf \Phi(\psi)=\textup{diag}\left\{ [1,e^{j\psi},\cdots,e^{j(M-1)\psi}]\right\}$
for $\psi\in[-\frac{\pi}{M},\frac{\pi}{M}]$.
In doing so, \eqref{equ:LS} is equivalent to $\min_{\bm \alpha,\{\theta_l\}^{L-1}_{l=0},\psi} \|\hat{\mathbf h}^{\text{ini}}_{k,p}-\mathbf \Phi^H(\psi)\mathbf F^H\bm \alpha\|^2$,
and then the angle estimation problem is boiled down to determining the optimal spatial rotation parameter $\psi_{k,p}$ and
the corresponding index set $\mathcal{Q}_{k,p}$ of nonzero components of $\bm \alpha$, namely,
\begin{align}\label{equ:phaserotationobjective}
  [\psi_{k,p},\mathcal{Q}_{k,p}]=\argmax_{\psi,\ \mathcal{Q}}\quad &\Big\|\left[\mathbf F\mathbf \Phi(\psi) \hat{\mathbf h}^{\text{ini}}_{k,p}\right]_{\mathcal{Q},:}\Big\|^2
  \quad \text{s.t.} \quad  |\mathcal{Q}|=\kappa,
\end{align}
where $\kappa$ is a pre-determined largest cardinality of $\mathcal{Q}$.
Afterwards, the AOA range of user-$k$ could be inferred from the indices inside $\mathcal{Q}_{k,p}$
according to Lemma 1 in \cite{tvt} as
\begin{align}\label{equ:A_hat}
  \hat{\mathcal{A}}_{k,p} =\left\{\vartheta\Big|\left(\frac{f_u d}{2\pi c}\psi_{k,p}-\frac{qc}{Mf_u d}\right)_{\min}\leq \cos(\vartheta) \leq\left(\frac{f_ud}{2\pi c}\psi_{k,p}-\frac{qc}{Mf_ud}\right)_{\max}, \forall q\in\mathcal{Q}_{k,p}\right\}.
\end{align}
Therefore, $\widehat{\bar\vartheta_{k,p}}$ and $\hat{\Delta}_{k,p}$ are obtained accordingly.

\subsubsection{Initial User Scheduling}
\begin{figure}[t]
\centering
\includegraphics[width=100mm]{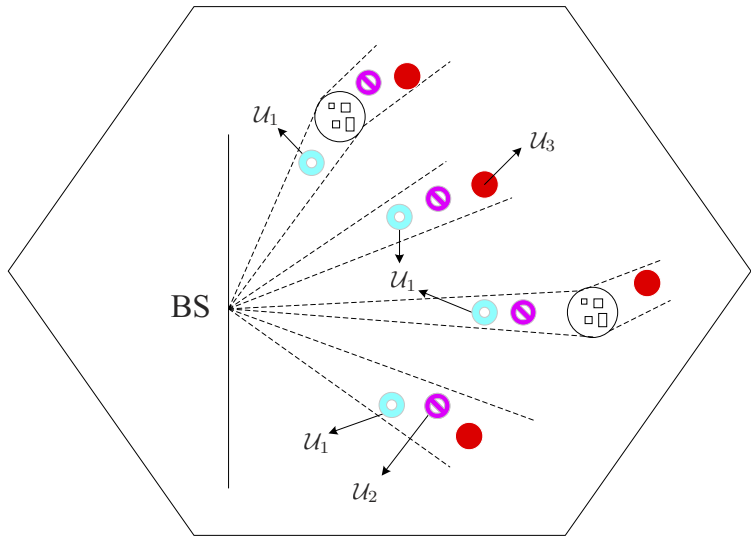}
\caption{ Illustration of ADMA user scheduling, where users (circles) with the same shape are in
the same group and can reuse the same orthogonal training sequence.
\label{fig:cov_userscheduling}}
\end{figure}

After obtaining the AOA intervals of all users, we divide
users into different groups, denoted by $\mathcal{U}_1,\mathcal{U}_2,\ldots$, such that users with non-overlapped
AOA intervals, i.e., $\hat{\mathcal{A}}_{k,p}$ or $\mathcal{Q}_{k,p}$, stay in the same group, as shown in
Fig. \ref{fig:cov_userscheduling}.
According to \cite{Caire,yin,tvt}, users in the same group could  reuse the same training sequence
in the subsequent channel updating without causing pilot contamination, while
different groups still adopt orthogonal training sequences to remove the inter-group interference.
Such an angle domain grouping strategy will significantly reduce the subsequent training overhead (after preamble) and is called as angle-division multiple access (ADMA).

\subsection{Uplink Channel Estimation with  CCM Reconstruction}
\subsubsection{Instantaneous Channel Update}
After ADMA user scheduling, the subsequent instantaneous channel update for all users could be
implemented with the spatial basis expansion model (SBEM) scheme proposed by \cite{tvt}, which resorts to the full utilization of the angular
information.
Take users in group $\mathcal{U}_1$ for instance and denote the orthogonal training sequence assigned to $\mathcal{U}_1$  as $\mathbf s_1$ with $\mathbf s_1^H\mathbf s_1=\rho_u$,
where $\rho_u$ is the total uplink training SNR.  Then at the $n$-th ($n=1,2,\ldots,N$) coherence time interval,
the received signals at BS from users in $\mathcal{U}_1$ are expressed as
\begin{align}
  \mathbf Y_{\mathcal{U}_1}(n) = \sum_{\substack{i\in\mathcal{U}_1}}\mathbf h_{i,p}(n)\mathbf s_1^H +\mathbf N,
\end{align}
where $\mathbf N$ is a Gaussian noise matrix with $\mathcal{CN}(0,1)$ elements. The inter-group interference does not appear due to the orthogonal training among different groups.
The uplink channel estimate for all users in $\mathcal{U}_1$ is then computed as
\begin{align}\label{equ:ULinstantChanl}
  \hat{\mathbf h}_{\mathcal{U}_1}(n) =\frac{1}{\rho_u}\mathbf Y_{\mathcal{U}_1}(n)\mathbf s_1=\mathbf h_{k,p}(n) + \sum_{\substack{i\neq k;i,k\in\mathcal{U}_1}}\mathbf h_{i,p}(n)+\frac{1}{\sqrt{\rho_u}}\mathbf n_{k,p},
\end{align}
with $\mathbf n_{k,p}\sim\mathcal{CN}(\mathbf 0,\mathbf I_M)$ denoting the normalized noise vector.
Since users in $\mathcal{U}_1$ have disjoint AOA intervals, the second term in \eqref{equ:ULinstantChanl} could be eliminated
 for the $k$-th user by only extracting the components of $\hat{\mathbf h}_{\mathcal{U}_1}(n)$ indexed by
 $\mathcal{Q}_{k,p}$. Specifically,  we have
\begin{align}\label{equ:ULinstantChanl2}
  & \left[\mathbf F\mathbf \Phi(\psi_{k,p}(n{-}1))\hat{\mathbf h}_{\mathcal{U}_1}(n)\right]_{\mathcal{Q}_{k,p}(n{-}1),:} =
  \left[\mathbf F\mathbf \Phi(\psi_{k,p}(n{-}1))\mathbf h_{k,p}(n)\right]_{\mathcal{Q}_{k,p}(n{-}1),:} \notag\\
  &\kern 40pt+\sum_{\substack{i\neq k;i,k\in\mathcal{U}_1}} \left[\mathbf F\mathbf \Phi(\psi_{k,p}(n{-}1))\mathbf h_{i,p}(n)\right]_{\mathcal{Q}_{k,p}(n{-}1),:}
  +\frac{1}{\sqrt{\rho_u}}\left[\mathbf F\mathbf \Phi(\psi_{k,p}(n{-}1))\mathbf n_{k,p}\right]_{\mathcal{Q}_{k,p}(n{-}1),:}\notag\\
  &\kern 40pt\approx\left[\mathbf F\mathbf \Phi(\psi_{k,p}(n{-}1))\mathbf h_{k,p}(n)\right]_{\mathcal{Q}_{k,p}(n{-}1),:} +
  \frac{1}{\sqrt{\rho_u}}\left[\mathbf F\mathbf \Phi(\psi_{k,p}(n{-}1))\mathbf n_{k,p}\right]_{\mathcal{Q}_{k,p}(n{-}1),:}.
\end{align}
Note that the parameters $\psi_{k,p}(n{-}1)$ and $\mathcal{Q}_{k,p}(n{-}1)$ from the
$(n{-}1)$-th coherence time are adopted for the current channel update, because each user's AOA interval is unlikely to change   much   in a short time slot. Then user-$k$'s uplink channel update could be
recovered from \eqref{equ:ULinstantChanl2} by the following equation:
\begin{align}\label{equ:ULinstantChannelLS}
\hat{\mathbf h}_{k,p}^{\text{SBEM}}(n)= \mathbf \Phi(\psi_{k,p}(n{-}1))^H\left[\mathbf F^H\right]_{:,\mathcal{Q}_{k,p}(n{-}1)}
  \left[\mathbf F\mathbf \Phi(\psi_{k,p}(n{-}1))\hat{\mathbf h}_{\mathcal{U}_1}(n)\right]_{\mathcal{Q}_{k,p}(n{-}1),:}.
\end{align}

\subsubsection{Instantaneous Angle Update}
With updated instantaneous channels, we are able to update the instantaneous $\hat{\mathcal{A}}_{k,p}(n)$,
$\mathcal{Q}_{k,p}(n)$ and $\psi_{k,p}(n)$ from $\hat{\mathbf h}^{\text{SBEM}}_{k,p}(n)$,
according to the same procedures of \eqref{equ:phaserotationobjective} in preamble phase.
This updated angle information actually helps to monitor the motion of users and thus
triggers a user re-scheduling process when two users' angular distance becomes smaller than a certain threshold,
as indicated by the flow chart in Fig. \ref{fig:cov_flowchart}.

\subsubsection{PAS Estimation}
As mentioned in Section II-B, we need to estimate the discrete PAS $S_{k,p}(\vartheta)$ for $\vartheta\in\hat{\mathcal{A}}_{k,p}(n)$.
Since only one instantaneous channel realization is obtained, we will first estimate  $\hat{\alpha}(\vartheta)$ from  $\hat{\mathbf h}^{\text{SBEM}}_{k,p}(n)$ and  then  use $|\hat{\alpha}(\vartheta)|^2$ to approximate the expectation $S_{k,p}(\vartheta)=\mathbb{E}\{|\alpha(\vartheta)|^2\}$.

The nonzero components of $\hat{\bm \alpha}$ in \eqref{equ:LS} could have been
used to estimate the desired $S_p(\theta_l)$ at their corresponding angles $\theta_l$.
However, the selected angles $\{\theta_l\}_{l=0}^{L-1}$ in \eqref{equ:LS} are chosen from
$[-\pi,\pi]$, not from $\hat{\mathcal{A}}_{k,p}(n)$. In other words, only few angles among $\{\theta_l\}_{l=0}^{L-1}$ will be located
inside $\hat{\mathcal{A}}_{k,p}(n)$ and thus the number of nonzero components of $\hat{\bm \alpha}$ in \eqref{equ:LS}
is not sufficient to get an accurate estimation of the continuous PAS $S_{k,p}(\vartheta)$ for $\vartheta\in\hat{\mathcal{A}}_{k,p}(n)$,
as indicated in \eqref{equ:discreteR}.
In line of this thought, we should first determine enough discrete angles of interest  inside $\hat{\mathcal{A}}_{k,p}(n)$ and then
estimate the corresponding instantaneous channel gains.
Specifically, with updated instantaneous channel $\hat{\mathbf h}^{\text{SBEM}}_{k,p}(n)$ as well as
estimated angle parameters $\hat{\mathcal{A}}_{k,p}(n)$, $\widehat{\bar{\vartheta}_{k,p}}(n)$ and $\hat{\Delta}_{k,p}(n)$,
the acquisition of instantaneous channel gains corresponding to selected angles inside $\hat{\mathcal{A}}_{k,p}(n)$
could be formulated as
\begin{align} \label{equ:AngleEstimation}
  \hat{\bm \alpha}(\bm \vartheta)=\argmin_{\bm \alpha(\bm \vartheta)}\quad \|\hat{\mathbf h}^{\text{SBEM}}_{k,p}(n) -\mathbf \Phi^H(\psi_{k,p}(n))\mathbf A(\bm \vartheta)\bm \alpha(\bm \vartheta)\|^2,
\end{align}
where $\bm \vartheta=[\vartheta_0,\vartheta_1,\ldots,\vartheta_{L-1}]^T$ collects $L$ discrete angles of interest selected from
$\hat{\mathcal{A}}_{k,p}(n)$. One simple way to determine these $L$ discrete angles is
to pick $L$ evenly spaced angles inside $\hat{\mathcal{A}}_{k,p}(n)$, namely, $\vartheta_l =\widehat{\bar{\vartheta}_{k,p}}(n)-\hat{\Delta}_{k,p}(n)+ \frac{2\hat{\Delta}_{k,p}(n) l}{L}$ for $l=0,\ldots,L-1$.
Moreover, $\mathbf A$ in \eqref{equ:LS} is replaced by $\mathbf A(\bm \vartheta)\triangleq[\mathbf a(\vartheta_0),\mathbf a(\vartheta_1),\ldots,\mathbf a(\vartheta_{L-1})]$, and
$\hat{\bm \alpha}(\bm \vartheta)=[\hat{\alpha}(\vartheta_0),\hat{\alpha}(\vartheta_1),\ldots,\hat{\alpha}(\vartheta_{L-1})]^T$
is exactly our desired instantaneous channel gains.

To solve the problem in \eqref{equ:AngleEstimation}, we have to consider two cases. When $L\leq M$,
$\mathbf A(\bm\vartheta)\in\mathbb{C}^{M\times L}$ is a tall matrix and the solution of $\hat{\bm \alpha}(\bm \vartheta)$
could be directly obtained by LS as
\begin{align}\label{equ:TnstanAlpha}
  \hat{\bm \alpha}(\bm \vartheta)=  \mathbf A^{\dag}(\bm \vartheta)\mathbf \Phi(\psi_{k,p}(n))\hat{\mathbf h}^{\text{SBEM}}_{k,p}(n).
\end{align}
On the other side,  a large $L~(\geq M)$ may generally  be considered to get dense angle samples inside $\hat{\mathcal{A}}_{k,p}(n)$
such that $|\hat{\alpha}(\vartheta)|^2$ may give a more accurate approximation of the continuous  $S_{k,p}(\vartheta)$.
Unfortunately,  due to more unknown variables than the observations,  LS method is not applicable for \eqref{equ:AngleEstimation}.
To find an alternative solution, let us consider \eqref{equ:AngleEstimation}
from another perspective. If we take the ideal channel vector $\mathbf h_{k,p}(n)$ as an $M$-point receiving vector in the spatial domain,
the values of $\bm \alpha(\bm \vartheta)$ are exactly the outputs of the discrete time Fourier transform (DTFT) of $\mathbf \Phi(\psi_{k,p}(n))\mathbf h_{k,p}(n)$ at certain angles. To be concrete, the DTFT of $\mathbf \Phi(\psi_{k,p}(n))\mathbf h_{k,p}(n)$ could be given as
\begin{align}\label{equ:betaxi}
  \beta(\xi) &\triangleq \frac{1}{\sqrt{M}}\sum_{m=0}^{M-1} [\mathbf \Phi(\psi_{k,p}(n))\mathbf h_{k,p}(n)]_m e^{-jm\xi}=\frac{1}{M}\int_{\vartheta\in\mathcal{A}_{k,p}(n)}\alpha(\vartheta)\sum_{m=0}^{M-1}e^{-j\chi m\cos(\vartheta)+jm\psi_{k,p}(n)-jm\xi}d\vartheta\notag\\
  &=\int_{\vartheta\in\mathcal{A}_{k,p}(n)}\alpha(\vartheta)~\frac{\sin\left[\left(\chi\cos(\vartheta)-\psi_{k,p}(n)+\xi\right)\frac{M}{2}\right]}
  {M\sin\left[\left(\chi\cos(\vartheta)-\psi_{k,p}(n)+\xi\right)\frac{1}{2}\right]}
  e^{-j\frac{M-1}{2}\left(\chi\cos(\vartheta)-\psi_{k,p}(n)+\xi\right)}d\vartheta\notag\\
  &=\int_{\vartheta\in\mathcal{A}_{k,p}(n)}\alpha(\vartheta)~\text{asinc}_M\left(\chi\cos(\vartheta)-\psi_{k,p}(n)+\xi\right)
  e^{-j\frac{M-1}{2}\left(\chi\cos(\vartheta)-\psi_{k,p}(n)+\xi\right)}d\vartheta,
\end{align}
where $\text{asinc}_M(x)\triangleq\frac{\sin(Mx/2)}{M\sin(x/2)}$ is an aliased sinc function,
$\beta(\xi)$ is the DTFT output and the scalar variable $\xi\in[-\pi,\pi]$ could be deemed as spatial frequency.
Since that $\text{asinc}_M(0)=1$ and $|\text{asinc}_M(x)|\leq \frac{1}{M|x|} $ for $x\neq 0$, there is
\begin{align}
  \beta (\xi)\big|_{\xi=\psi_{k,p}(n)-\chi\cos(\vartheta)} \to \alpha(\vartheta), \ \forall \vartheta \in \mathcal{A}_{k,p}(n),\ \text{as}\ M\to\infty.
\end{align}
which indicates that as $\xi$ varies across the whole spatial frequency range, $\beta(\xi)$ gives an accurate estimation for $\alpha(\vartheta),~\forall \vartheta\in\mathcal{A}_{k,p}(n)$. When $L$ discrete angles
$\bm \vartheta=[\vartheta_0,\vartheta_1,\ldots,\vartheta_{L-1}]^T$ is considered, the desired
$\hat{\bm \alpha}(\bm \vartheta)=[\hat{\alpha}(\vartheta_0),\hat{\alpha}(\vartheta_1),\ldots,\hat{\alpha}(\vartheta_{L-1})]^T$
are exactly the sampling results of $\beta (\xi)$ at corresponding spatial frequencies, namely,
$\bm \beta(\bm \xi)\triangleq[\beta(\xi_0),\beta(\xi_1),\ldots,\beta(\xi_{L-1})]^T$, with
$\xi_l=\psi_{k,p}(n)-\chi\cos(\vartheta_l)$ for $l=0,\ldots,L-1$. According to \eqref{equ:betaxi},
the expression of $\bm \beta(\bm \xi)$ could be further specified as
\begin{align}
  \bm\beta(\bm \xi) = \bm \Psi \mathbf \Phi(\psi_{k,p}(n))\mathbf h_{k,p}(n),
\end{align}
where $\bm \Psi\in\mathbb{C}^{L\times M}$ is defined as
\begin{align}
\bm \Psi \triangleq \left[
  \begin{array}{cccc}
   1 &  e^{-j\xi_0}  & \cdots & e^{-j(M-1)\xi_0} \\
   1 &  e^{-j\xi_1}  & \cdots & e^{-j(M-1)\xi_1} \\
    \vdots & \vdots & \ddots & \vdots\\
   1 &  e^{-j\xi_{L-1}}  & \cdots & e^{-j(M-1)\xi_{L-1}}
\end{array}\right].
\end{align}
Therefore, the instantaneous channel gain estimates $\hat{\bm \alpha}(\bm \vartheta)$ under the
condition of $L\geq M$ can be expressed as
\begin{align}\label{equ:30}
  \hat{\bm \alpha}(\bm \vartheta)=  \bm \Psi\mathbf \Phi(\psi_{k,p}(n))\hat{\mathbf h}^{\text{SBEM}}_{k,p}(n).
\end{align}

\subsubsection{Uplink CCM Reconstruction}
In the sequel, we show how to reconstruct the uplink CCMs with the obtained uplink angles $\hat{\mathcal{A}}_{k,p}(n)$ and channel gains $\hat{\bm \alpha}(\bm \vartheta)$.
%
Let us model the random channel phases of all signal rays as independent uniformly distributed variables in $[-\pi,\pi]$,
and then along with the estimated channel gains, we may artificially generate many copies of instantaneous uplink channel estimates. In particular, we generate the auxiliary uplink channel estimates with random additional phase as
\begin{align}\label{equ:aux}
   \mathbf h_{k,p}^{\text{aux}}(n) =\int_{\hat{\mathcal{A}}_{k,p}(n)}|\hat{\alpha}(\vartheta)|e^{j\phi(\vartheta)+j\widetilde{\phi}(\vartheta)}\mathbf a(\vartheta)d\vartheta,
\end{align}
where $\widetilde{\phi}(\vartheta)$ is the modeled random phase for the incident ray from AOA $\vartheta$, and
$ \widetilde{\phi}(\vartheta_i)$ and $\widetilde{\phi}(\vartheta_j)$ are mutually independent for any $\vartheta_i,\vartheta_j\in\hat{\mathcal{A}}_{k,p}(n)$ and $\vartheta_i\neq \vartheta_j$. Approximating the integral of \eqref{equ:aux} with the discrete angles $\bm \vartheta$ and the estimated channel gains $\hat{\bm \alpha}(\bm \vartheta)$ in \eqref{equ:AngleEstimation}, we obtain
\begin{align}
   \mathbf h_{k,p}^{\text{aux}}(n)&\approx \sum_{l=0}^{L-1}\hat{\alpha}(\vartheta_l)e^{j\widetilde{\phi}(\vartheta_l)}\mathbf \Phi^H(\psi_{k,p}(n))\mathbf a(\vartheta_l)
   =\mathbf \Phi^H(\psi_{k,p}(n))\mathbf A(\bm \vartheta)\mathbf \Pi(\bm \vartheta)\hat{\bm \alpha}(\bm \vartheta)\notag\\
   &=\mathbf \Phi^H(\psi_{k,p}(n))\mathbf A(\bm \vartheta)\mathbf \Pi(\bm \vartheta)\bm\Psi\mathbf \Phi(\psi_{k,p}(n))\hat{\mathbf h}^{\text{SBEM}}_{k,p}(n),
\end{align}
where $\mathbf \Pi(\bm \vartheta)\triangleq\text{diag}\big\{\widetilde{\phi}(\vartheta_0),\widetilde{\phi}(\vartheta_1),\ldots,\widetilde{\phi}(\vartheta_{L-1})\big\}$
collects the random phases. To reconstruct the CCM $\hat{\mathbf R}^u_{k,p}(n)$, we take the expectation of $\mathbf h_{k,p}^{\text{aux}}(n)$ and obtain
\begin{align}
  \hat{\mathbf R}^u_{k,p}(n) &= \mathbb{E}\left\{\mathbf h_{k,p}^{\text{aux}}(n)\mathbf h_{k,p}^{\text{aux}}(n)^H\right\}\notag\\
  &=\mathbb{E}\left\{\mathbf \Phi^H(\psi_{k,p}(n))\mathbf A(\bm \vartheta)\mathbf \Pi(\bm \vartheta)\hat{\bm \alpha}(\bm \vartheta)\hat{\bm \alpha}^H(\bm \vartheta)\mathbf \Pi^H(\bm \vartheta)\mathbf A^H(\bm \vartheta)\mathbf \Phi(\psi_{k,p}(n))\right\}\notag\\
  &=\mathbf \Phi^H(\psi_{k,p}(n))\mathbf A(\bm \vartheta)\mathbb{E}\left\{\left(\hat{\bm \alpha}(\bm \vartheta)\hat{\bm \alpha}^H(\bm \vartheta)\right)\odot \mathbf T  \right\}\mathbf A^H(\bm \vartheta)\mathbf \Phi(\psi_{k,p}(n))\notag\\
  &=\mathbf \Phi^H(\psi_{k,p}(n))\mathbf A(\bm \vartheta)\left( \left(\hat{\bm \alpha}(\bm \vartheta)\hat{\bm \alpha}^H(\bm \vartheta)\right)\odot \mathbb{E}\{\mathbf T\} \right)\mathbf A^H(\bm \vartheta)\mathbf \Phi(\psi_{k,p}(n)),
\end{align}
where $\odot$ denotes the Hadamard product (elementwise product) and $\mathbf T$ is defined as
\begin{align}
\mathbf T\triangleq\left[
  \begin{array}{cccc}
   1 &  e^{j(\widetilde{\phi}(\vartheta_0)-\widetilde{\phi}(\vartheta_1))}  & \cdots & e^{j(\widetilde{\phi}(\vartheta_0)-\widetilde{\phi}(\vartheta_{L-1}))} \\
    e^{j(\widetilde{\phi}(\vartheta_1)-\widetilde{\phi}(\vartheta_0))} & 1& \cdots & e^{j(\widetilde{\phi}(\vartheta_1)-\widetilde{\phi}(\vartheta_{L-1}))} \\
    \vdots & \vdots & \ddots & \vdots\\
    e^{j(\widetilde{\phi}(\vartheta_{L-1})-\widetilde{\phi}(\vartheta_{0}))} & e^{j(\widetilde{\phi}(\vartheta_{L-1})-\widetilde{\phi}(\vartheta_{1}))}& \cdots & 1\\
\end{array}\right].
\end{align}
Since that $ \widetilde{\phi}(\vartheta_i)$ and $\widetilde{\phi}(\vartheta_j)$ are mutually independent, it leads to $\mathbf E\{e^{j(\widetilde{\phi}(\vartheta_i)-\widetilde{\phi}(\vartheta_j))}\}=\delta(\vartheta_i-\vartheta_j)$ and $\mathbb{E}\{\mathbf T\}=\mathbf I_L$, and thereof
\begin{align}\label{equ:ULCCM}
  \hat{\mathbf R}_{k,p}^u(n) &= \mathbf \Phi^H(\psi_{k,p}(n))\mathbf A(\bm \vartheta)\underbrace{\text{diag}\left\{|\hat{\alpha}(\vartheta_0)|^2,|\hat{\alpha}(\vartheta_1)|^2,\ldots,|\hat{\alpha}(\vartheta_{L-1})|^2\right\}}_{\text{estimated PAS}\ \{S_p^u(\vartheta_0),S_p^u(\vartheta_1),\ldots,S_p^u(\vartheta_{L-1})\}}\mathbf A^H(\bm \vartheta)\mathbf \Phi(\psi_{k,p}(n))\notag\\
  &=\sum_{l=0, \vartheta_l\in\bm \vartheta}^{L-1}|\hat{\alpha}(\vartheta_l)|^2\left(\mathbf{\Phi}^H(\psi_{k,p}(n))\mathbf a_u(\vartheta_l)\mathbf a_u^H(\vartheta_l)\mathbf \Phi(\psi_{k,p}(n))\right).
\end{align}
From (\ref{equ:ULCCM}), it tells that the channel covariance is mainly related with the
angular information and PAS but is not related to the channel phase information, which matches the intuition and
the expression (4) quite well.

The main steps of above CCM reconstruction scheme is summarized in Algorithm 1.
Considering the characteristics of the proposed CCM reconstruction scheme in this section,
we will refer to it as \emph{Instantaneous Channel aided parametric CCM reconstruction (IC-pCCM)},
for distinguishing from the \emph{MC-iCCM} and \emph{CF-iCCM} schemes in section II-B.

\begin{algorithm}[t]
\caption{: Main steps of proposed \emph{IC-pCCM} scheme.}
\label{alg::userscheduling}
\begin{itemize}
\item \textbf{Step 1:} After ADMA user scheduling, obtain $\hat{\mathbf h}_{k,p}^{\text{SBEM}}(n)$ via \eqref{equ:ULinstantChannelLS} for each user,
and update $\hat{\mathcal{A}}_{k,p}(n)$, $\mathcal{Q}_{k,p}(n)$ and $\psi_{k,p}(n)$ via \eqref{equ:phaserotationobjective} and \eqref{equ:A_hat}.
\item \textbf{Step 2:} Determine $L$ discrete angles of interest within $\hat{\mathcal{A}}_{k,p}(n)$, namely, $\bm \vartheta=[\vartheta_0,\vartheta_1,\ldots,\vartheta_{L-1}]^T$ with $\vartheta_l\in\hat{\mathcal{A}}_{k,p}(n)$.
\item \textbf{Step 3:} Estimate the instantaneous channel gain $\hat{\bm \alpha}(\bm \vartheta)$ corresponding to these discrete angles $\bm \vartheta$
from $\hat{\mathbf h}_{k,p}^{\text{SBEM}}(n)$, via \eqref{equ:TnstanAlpha} if $L\leq M$ or via \eqref{equ:30} if $L>M$.
\item \textbf{Step 4:} Reconstruct the uplink CCM via \eqref{equ:ULCCM}.
\end{itemize}
\end{algorithm}

\subsubsection{Improved Uplink Channel Estimation With Reconstructed CCMs}
With the reconstructed CCMs in \eqref{equ:ULCCM}, the uplink channel estimation performances could be further
improved by applying MMSE estimator for \eqref{equ:ULinstantChannelLS}, namely,
\begin{align}\label{equ:ULmmse}
    \hat{\mathbf h}^{\text{mmse}}_{k,p}(n) &= \hat{\mathbf R}_{k,p}^u(n)\Big(\frac{1}{\rho_u}\mathbf I_M+\sum_{i\in\mathcal{U}_1}\hat{\mathbf R}_{i,p}^u(n)\Big)^{-1}\hat{\mathbf h}^{\text{SBEM}}_{k,p}(n)\notag\\
    &=\mathbf V^u_{k,p}\mathbf \Sigma^u_{k,p}{\mathbf V^u_{k,p}}^H\Big(\frac{1}{\rho_u}\mathbf I_M+\sum_{i\in\mathcal{U}_1}\mathbf V^u_{i,p}\mathbf \Sigma^u_{i,p}{\mathbf V^u_{i,p}}^H\Big)^{-1}\hat{\mathbf h}^{\text{SBEM}}_{k,p}(n)\notag\\
    &\longrightarrow \mathbf V^u_{k,p}\mathbf \Sigma^u_{k,p}\Big(\frac{1}{\rho_u}\mathbf I_M+\mathbf \Sigma^u_{k,p}\Big)^{-1}{\mathbf V^u_{k,p}}^H \hat{\mathbf h}^{\text{SBEM}}_{k,p}(n), \quad \text{as}\ M\to\infty,
\end{align}
where $\hat{\mathbf R}_{k,p}^u(n)$ is decomposed into $\hat{\mathbf R}_{k,p}^u(n) =\mathbf V^u_{k,p}\mathbf \Sigma^u_{k,p}{\mathbf V^u_{k,p}}^H$, with $\mathbf V^u_{k,p}$ being the signal eigenvector matrix of size $M\times \nu$, $\mathbf \Sigma^u_{k,p}$ being the eigenvalue matrix of $\nu \times \nu $
and $\nu\triangleq\text{rank}(\hat{\mathbf R}_{k,p}^u(n))$.
Due to the narrow AS condition, the reconstructed CCMs of all users would exhibit low-rank property \cite{Caire,yin,BDMA}, i.e., $\nu\ll M$, and thus the $\nu$ dominant eigenvectors of $\hat{\mathbf R}_{k,p}^u(n)$ are sufficient for
uplink training in \eqref{equ:ULmmse}. 
The asymptotic orthogonality between $\mathbf V^u_{k,p}$ and $\mathbf V^u_{i,p}$ is considered when $M\to\infty$ \cite{Caire,yin}.
It is expected from \eqref{equ:ULmmse} that the reconstructed CCMs could  enhance the performances of
uplink channel estimation compared to the SBEM method in \eqref{equ:ULinstantChannelLS}.
It need to be mentioned that this channel estimation performance enhancement
as well as above CCM construction are both obtained without any additional training overhead.

\subsection{Downlink CCM Inference and Downlink Channel Estimation}
\subsubsection{Downlink CCM Inference}
According to the derivation of downlink CCM inference from uplink measurements in \eqref{equ:ULDLtransdiscrete}, together with
the uplink angle and PAS estimates, the downlink CCM $\hat{\mathbf R}_{k,p}^d(n)$ in the $n$-th coherence time could be inferred as
\begin{align} \label{equ:DLCCM}
  \hat{\mathbf R}_{k,p}^d(n) =\sum_{l=0, \vartheta_l\in\bm \vartheta}^{L-1}\mu |\hat{\alpha}(\vartheta_l)|^2\mathbf\Theta(\vartheta_l)\mathbf{\Phi}^H(\psi_{k,p}(n))\mathbf a_u(\vartheta_l)\mathbf a_u^H(\vartheta_l)\mathbf \Phi(\psi_{k,p}(n))\mathbf \Theta^H(\vartheta_l).
\end{align}
As mentioned in section II-C, there maybe exist an unknown amplitude scaling factor $\mu$
between the estimated and the real downlink CCMs. However, to be noted that this
unknown factor does not affect the eigenvectors, or say, the signal subspace, and thus
is not going to impact the subsequent eigen-beamforming for donwlink  channel estimation.


\subsubsection{Downlink Training With Eigen-Beamforming}
With the inferred downlink CCM at hand, the optimal eigen-beamforming \cite{Caire,yin}
may be adopted for downlink CSI training.
Similar to the uplink case, BS only need to train along $\nu$ dominant eigenvectors of the low-rank $\hat{\mathbf R}_{k,p}^d(n)$
\cite{Caire,yin,BDMA}. This operation  could significantly
reduce the overall downlink training overhead, compared to the overhead of $M\times M$ orthogonal training matrix
required in conventional orthogonal downlink training scheme.
Denote the corresponding eigen-beamforming matrix as $\mathbf B_{k,p}$ for each user.
Then, the downlink training signals received at user-$k$ of group $\mathcal{U}_1$ is given as
\begin{align}
  \mathbf y_{k,p}^H(n) = {\mathbf h_{k,p}^d(n)}^H\Big(\sum_{i \in\mathcal{U}_1}\mathbf B_{i,p}\Big)\mathbf S+\mathbf n_{k,p}^H,
\end{align}
where $\mathbf n_{k,p}\sim\mathcal{CN}(\mathbf 0,\mathbf I_{\nu})$ and $\mathbf S$ is a $\nu\times\nu$ scaled unitary training matrix, i.e., $\mathbf S\mathbf S^H= \rho_d \mathbf I_{\nu}$ with
$ \rho_d$ being the downlink training power.
To complete the downlink training procedure, user-$k$ has to feed back the $\nu$ received signals $\mathbf y_{k,p}^H(n)$ to BS, and then
an MMSE estimator is adopted by BS to recover the downlink channel $\mathbf h_{k,p}^d(n)$ for user-$k$ as:
\begin{align}\label{equ:dlmmse}
\hat{\mathbf h}_{k,p}^d(n)
=\ &\hat{\mathbf R}_{k,p}^d(n)\Big(\sum_{i \in\mathcal{U}_1}\mathbf B_{i,p}\Big) \left(\sum_{i,j \in\mathcal{U}_1} \mathbf B^H_{i,p} \hat{\mathbf R}_{k,p}^d(n) \mathbf B_{j,p} + \frac{1}{\rho_d }\mathbf I_{\nu}\right)^{-1}\left(\Big(\sum_{i\in\mathcal{U}_1}\mathbf B_{i,p}^H\Big)\mathbf h_{k,p}^d(n)+\frac{1}{\rho_d }\mathbf S\mathbf n_{k,p}\right)\notag\\
\longrightarrow\ &\hat{\mathbf R}_{k,p}^d(n)\mathbf B_{k,p} \left(\mathbf B^H_{k,p}\hat{\mathbf R}_{k,p}^d(n)\mathbf B_{k,p} +\frac{1}{\rho_d }\mathbf I_{\nu}\right)^{-1}\left(\mathbf B_{k,p}^H\mathbf h_{k,p}^d(n)+\frac{1}{\rho_d }\mathbf S\mathbf n_{k,p}\right), \quad \text{as}\ M\to\infty.
\end{align}
Similarly, when $M\to\infty$, the asymptotic orthogonality suppresses the inter-user interference and promotes a better performance of
downlink training \cite{Caire,yin}.

\section{Simulations}
In this section, we demonstrate the effectiveness of the proposed strategy through numerical examples.
The system parameters are chosen as $M=128$, $f_u=2$ GHz, $d=\frac{c}{2f_u}$ and  $f_d-f_u=100$ MHz, unless otherwise mentioned.
Suppose users are randomly distributed in the coverage of the BS
and are dynamically grouped and scheduled for transmission as per their central AOA and ASs.
Without loss of generality, we consider one single MPC for all users and thus the channel models boil down to block flat fading,
while  the extension to multi-path channel model is straightforward.
The channel vectors of different users are formulated  according to \eqref{equ:channelmodel} with
different ASs  and PAS distributions, including uniform and Laplacian distributions.

\begin{figure}[t]
\centering
\includegraphics[width=120mm]{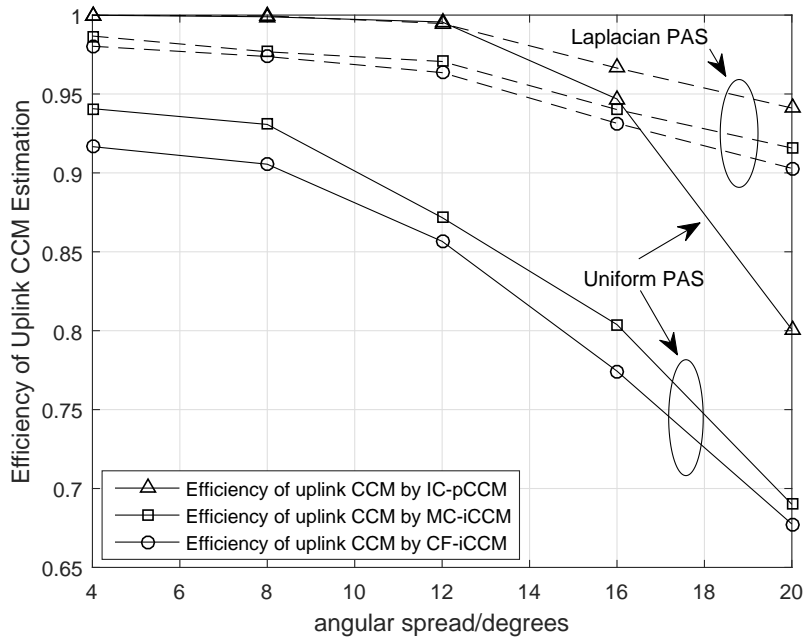}
\caption{ Efficiency of uplink CCMs reconstructed by three methods, with $\nu=16$ and SNR = $10$ dB. }
\label{fig:ULefficiency}
\end{figure}

\begin{figure}[t]
\centering
\includegraphics[width=120mm]{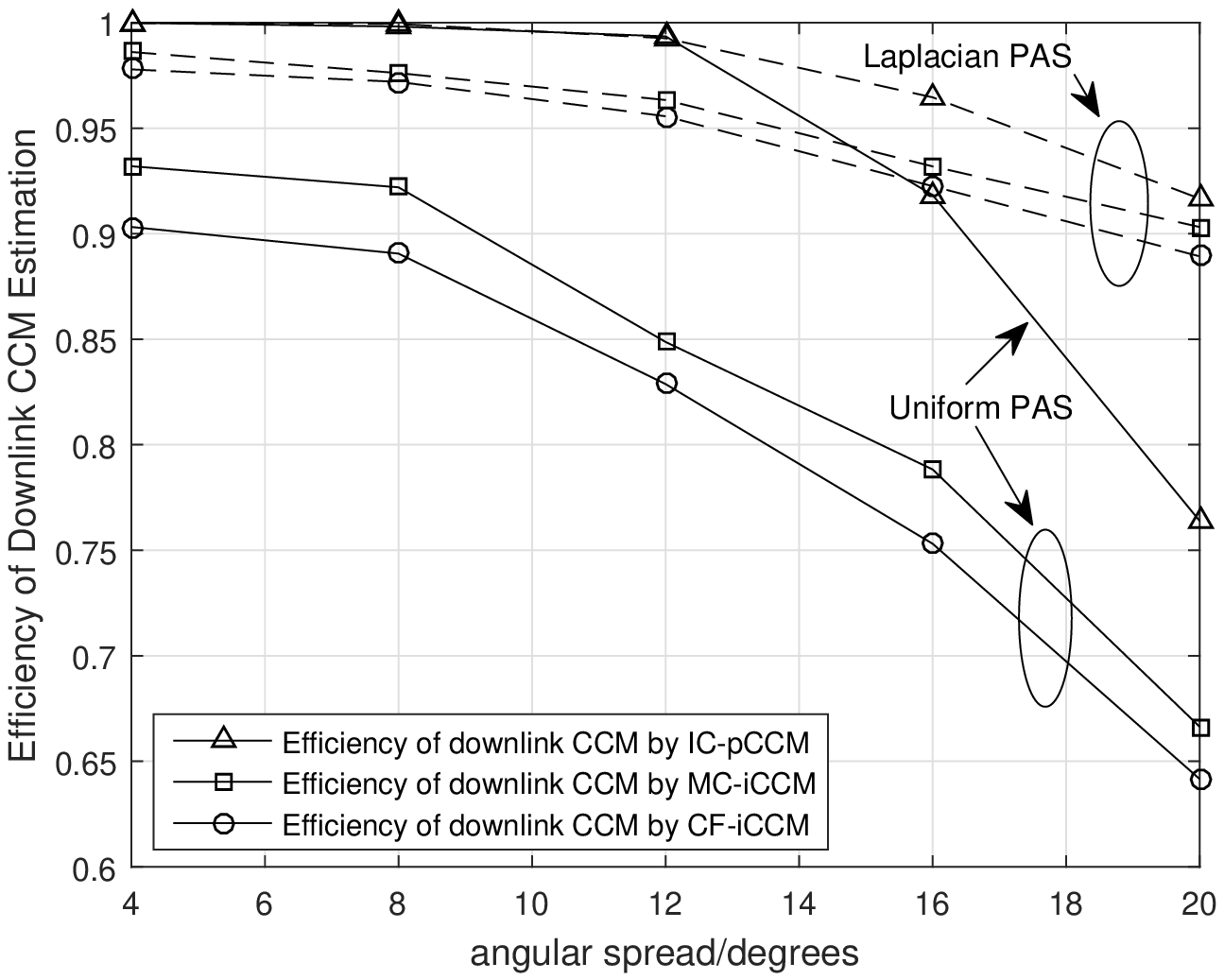}
\caption{ Efficiency of downlink CCMs reconstructed by three methods, with $\nu=16$ and SNR = $10$ dB. }
\label{fig:DLefficiency}
\end{figure}

In the first examples, we examine the  efficiency of uplink/downlink CCMs reconstruction, defined as
\begin{align}
  \eta\triangleq \frac{1}{N}\sum_{n=1}^N\frac{\langle\mathbf R_{k,p}(n),\hat{\mathbf B}_{k,p}\hat{\mathbf B}_{k,p}^H\rangle}{\text{tr}\left\{\mathbf R_{k,p}(n)\right\}}=\frac{1}{N}\sum_{n=1}^N\frac{\text{tr}\{\hat{\mathbf B}_{k,p}^H\mathbf R_{k,p}(n)\hat{\mathbf B}_{k,p}\}}{\text{tr}\left\{\mathbf R_{k,p}(n)\right\}},
\end{align}
 where $\mathbf R_{k,p}(n)$ denotes the real CCMs of user-$k$, namely, $\mathbf R_{k,p}^u(n)$ or $\mathbf R_{k,p}^d(n)$, and
 $\hat{\mathbf B}_{k,p}\in\mathbb{C}^{M\times \nu}$ includes the $\nu$ dominant eigenvectors of $\hat{\mathbf R}^u_{k,p}(n)$ or $\hat{\mathbf R}^d_{k,p}(n)$
corresponding to the $\nu$ largest eigenvalues. 
Obviously, for any given $\nu\ll M$, if $\eta\approx 1$, then a significant amount of received
signal's power is captured by a low-dimensional signal subspace, which also serves as an indicator of the ``similarity'' between estimated signal subspaces and the real ones.   Besides the \emph{IC-pCCM }scheme, the efficiencies of CCMs constructed by
\emph{CF-iCCM} and \emph{MC-iCCM} methods in section II-B with prior knowledge of PAS distributions  are also included in Fig. \ref{fig:ULefficiency} and Fig. \ref{fig:DLefficiency}.
For all the methods, the central AOA and AS parameters are estimated  from uplink instantaneous channel estimates
and then applied for uplink/downlink CCM reconstructions. Meanwhile,
the uplink/downlink training power is set as $\rho_u=\rho_d=10$ dB and
the default cardinality of $\mathcal{Q}_{k,p}$ for each user is set as $\kappa=\nu=16$.
It can be seen from Fig. \ref{fig:ULefficiency} and Fig. \ref{fig:DLefficiency} that when $\nu=16, M=128$, the \emph{IC-pCCM} scheme
could obtain a relative high efficiency for both uplink/downlink CCM estimation, for
both uniform and Laplacian PAS distributions, and for both narrow and wide ASs.
This demonstrates the effectiveness of CCM reconstruction
from instantaneous channel estimates.
Meanwhile,   the \emph{IC-pCCM} scheme outperforms the
\emph{CF-iCCM} and \emph{MC-iCCM} methods for both uplink and downlink CCMs construction.
Especially, the performances of \emph{CF-iCCM} method deteriorate significantly for larger AS and uniform PAS.
This is because the performances of \emph{CF-iCCM} rely heavily on the accuracy of angle estimation, which unfortunately is
often deteriorated when multiple users are scheduled at the same time. By contrast, the \emph{IC-pCCM} scheme is more robust
to the angle estimation error for the reason that the value of $\hat{\alpha}(\vartheta)$ will be small if an estimated $\vartheta$ does not
belong to the real AOA interval.
Furthermore, if Laplacian PAS is considered, the CCM reconstruction efficiencies of all three methods
increase for the reason that more channel power is concentrated in central DOAs.

\begin{figure}[t]
\centering
\includegraphics[width=120mm]{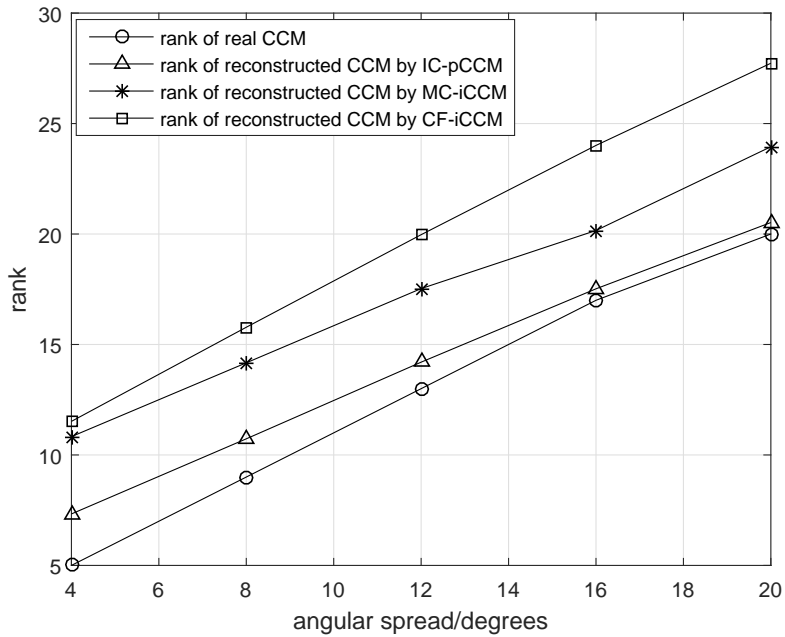}
\caption{ Comparison of Ranks of CCMs reconstructed by different methods, with uniform PAS and SNR=$10$ dB.
\label{fig:ULrank}}
\end{figure}

Fig. \ref{fig:ULrank}   compares the ranks of real uplink CCMs and the estimated uplink CCMs by the three methods
as a function of AS with
uniform PAS and SNR = $10$ dB. It shows that the ranks of CCMs estimated by \emph{IC-pCCM} is  close to the ranks of
real CCMs, while the ranks of CCMs reconstructed by \emph{CF-iCCM} and \emph{MC-iCCM} are much larger, which means that a larger angle estimation error
will cause a wider spread of the eigenvalues of estimated CCMs by integral methods.
Meanwhile, with increasing AS, the gap between ranks of
constructed CCMs by \emph{CF-iCCM} and \emph{MC-iCCM} also becomes larger. This verifies the discussion in section II-B that
\emph{MC-iCCM} is more accurate than \emph{CF-iCCM} for large AS.

\begin{figure}[t]
\centering
\includegraphics[width=120mm]{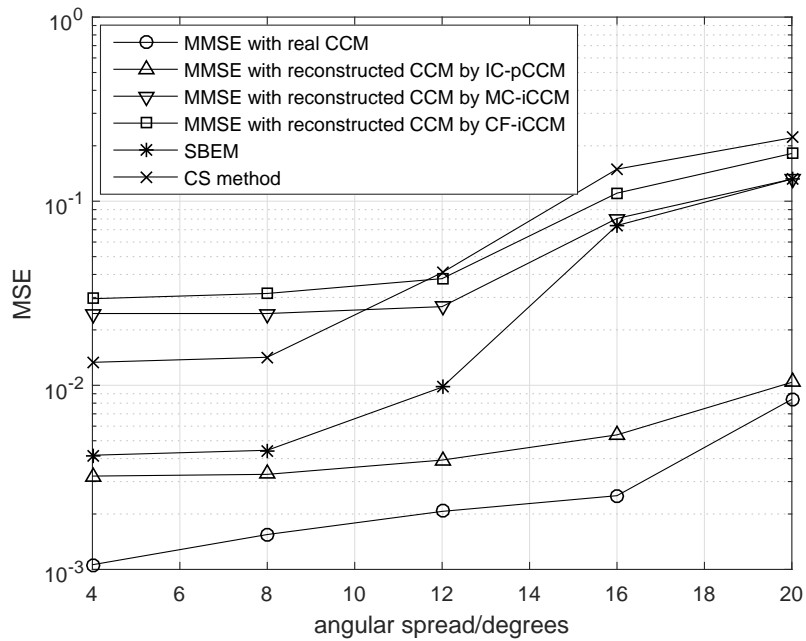}
\caption{ Uplink MSE performances of MMSE estimators with reconstructed CCMs as a function of AS, with $\nu=20$, SNR=$10$ dB and uniform PAS.
\label{fig:ULMSE_AS}}
\end{figure}

Fig. \ref{fig:ULMSE_AS} displays the performances of uplink channel estimation via MMSE estimator with CCMs reconstructed
by the three different methods. For better comparison, the SBEM
\cite{tvt} and the CS-based method in \cite{DaiTSP} are also included. 
The performance metric of the channel estimation is taken as the average individual MSE, i.e.,
\begin{equation*}
 \text{MSE} \triangleq \frac{1}{NK}\sum_{n=1}^N\sum_{k=1}^K\frac{\left\|\mathbf h_{k,p}(n)-\hat{\mathbf h}_{k,p}(n)\right\|^2}{\left\|\mathbf h_{k,p}(n)\right\|^2}.
\end{equation*}
For uplink training, the SNR is set as $\rho_u=10$ dB and only  $\nu=20$ dominant eigenvectors along with
the corresponding eigenvalues are used for uplink channel estimation in \eqref{equ:ULmmse}
for the consideration of limited radio frequency chains and the purpose of training overhead reduction.
Fig. \ref{fig:ULMSE_AS} illustrates that the MMSE estimator with CCMs reconstructed from uplink channel estimates by \emph{IC-pCCM}
scheme can achieve a significant performance improvement than that of SBEM and only has a small gap between the
performance of ideal CCM case. 
Meanwhile, as AS increases, the performances of SBEM, \emph{CF-iCCM}, \emph{MC-iCCM} and CS methods deteriorate greatly, while the MMSE performances
with real CCMs and reconstructed CCM via \emph{IC-pCCM} are relatively more robust to larger ASs. The reason lies in that
the eigenvectors of CCMs formulate  an optimal set of expansion basis vectors for signal expansion in the logical domain, such that
more signal power is accumulated in lower-dimensional signal subspace even in the larger AS conditions.
However, the CCMs estimated by \emph{CF-iCCM} and \emph{MC-iCCM} methods fail for the inaccurate angle estimation as explained above.

\begin{figure}[t]
\centering
\includegraphics[width=120mm]{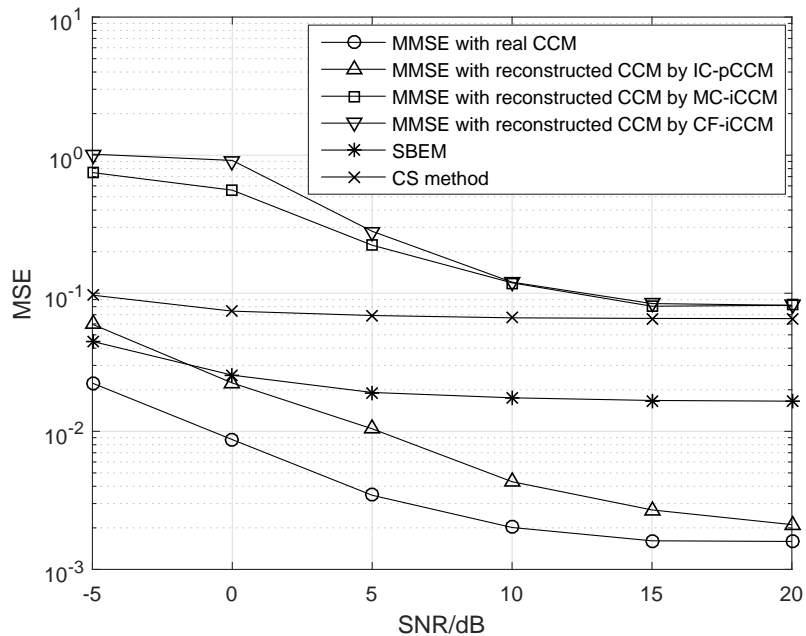}
\caption{Uplink MSE performances of MMSE estimators with reconstructed CCMs as a function of SNR,
with AS $\Delta_{k,p}=10^\circ$, $\nu=16$ and uniform PAS.
\label{fig:ULMSE_SNR}}
\end{figure}

Fig. \ref{fig:ULMSE_SNR} then compares the uplink MSE performances of the MMSE estimators
with reconstructed CCMs by different methods as well as the SBEM and CS methods as a function of SNR, with AS $\Delta_{k,p}=10^\circ$,
$\nu=16$ and a uniform PAS.
It can be seen that the uplink MSE curves of all methods  have their own error floors. There are two main reasons
for these error floors. First, only limited number of expansion basis vectors,
i.e., $\nu=16$ in Fig. \ref{fig:ULMSE_SNR}, is used for uplink training enhancement in \eqref{equ:ULmmse}. More importantly,
the instantaneous channel estimates in \eqref{equ:ULinstantChanl}-\eqref{equ:ULinstantChannelLS} have
their own truncation errors by only extracting components of $\hat{\mathbf h}_{\mathcal{U}_1}(n)$ indexed by $\mathcal{Q}_{k,p}$.
This truncation error cannot be improved even by the proposed uplink training enhancement scheme
and thus even the MSE curve with real CCM has its own error floor.
However, the MSE performance of MMSE with reconstructed CCM by \emph{IC-pCCM} method
has a significant improvement. As explained above, this is because the real CCMs as well as the reconstructed CCMs with
\emph{IC-pCCM} are able to accumulate more power with only $\nu=16$ eigenvectors, which also
corroborates the effectiveness and satisfying performance gains provided by \emph{IC-pCCM} scheme.

\begin{figure}[t]
\centering
\includegraphics[width=120mm]{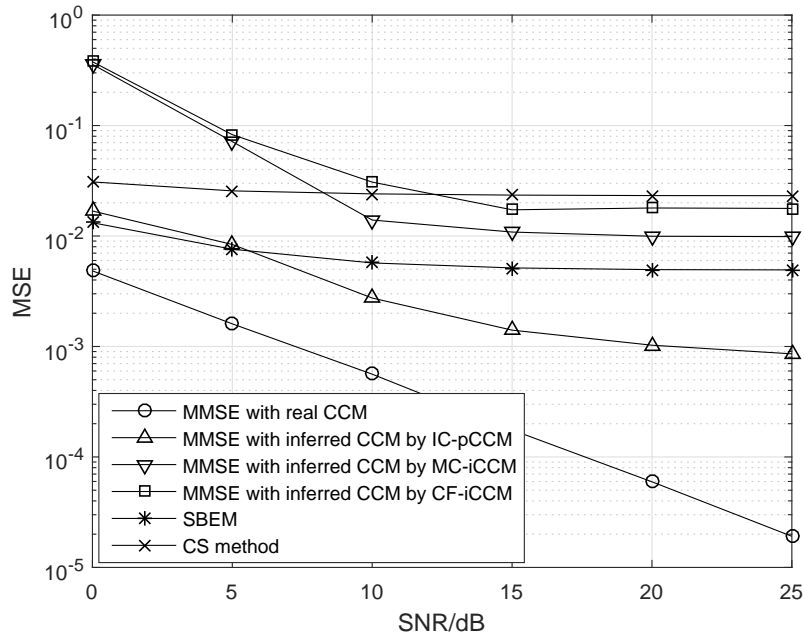}
\caption{ Downlink MSE performances of MMSE estimators with inferred CCMs as a function of SNR,
with AS $\Delta_{k,p}=10^\circ$, $\nu=20$ and uniform PAS.
\label{fig:DMLSE_SNR}}
\end{figure}

Fig. \ref{fig:DMLSE_SNR} illustrates the downlink MSE performances of the MMSE estimators
with inferred CCMs by different methods as well as the SBEM and CS methods as a function of SNR, with AS $\Delta_{k,p}=10^\circ$.
For donwlink training, only  $\nu=20$ dominant eigenvectors along with
the corresponding eigenvalues are used for eigen-beamforming in \eqref{equ:dlmmse}
for the same consideration of limited radio frequency chains and reduced training and feedback overhead.
It can be seen that the MSE curves of SBEM, \emph{CF-iCCM} and \emph{MC-iCCM} methods  have their own error floors.
Note that different from uplink case, these error floors are only due to
the limited number of expansion basis vectors for downlink training. Without truncation error, the MSE curve with
real CCM does not have any error floor.
Compared to \emph{CF-iCCM} and \emph{MC-iCCM}, the MSE performance of MMSE with inferred CCM by \emph{IC-pCCM} has been improved greatly.
Meanwhile, the gap between the MSE performances of \emph{CF-iCCM} and \emph{MC-iCCM} methods shows that
when AS is large or the narrow AS condition is not satisfied, the \emph{CF-iCCM} method \eqref{equ:6b} using Taylor expansion to
approximate \eqref{equ:6a} certainly results in performance loss.

\begin{figure}[t]
\centering
\includegraphics[width=120mm]{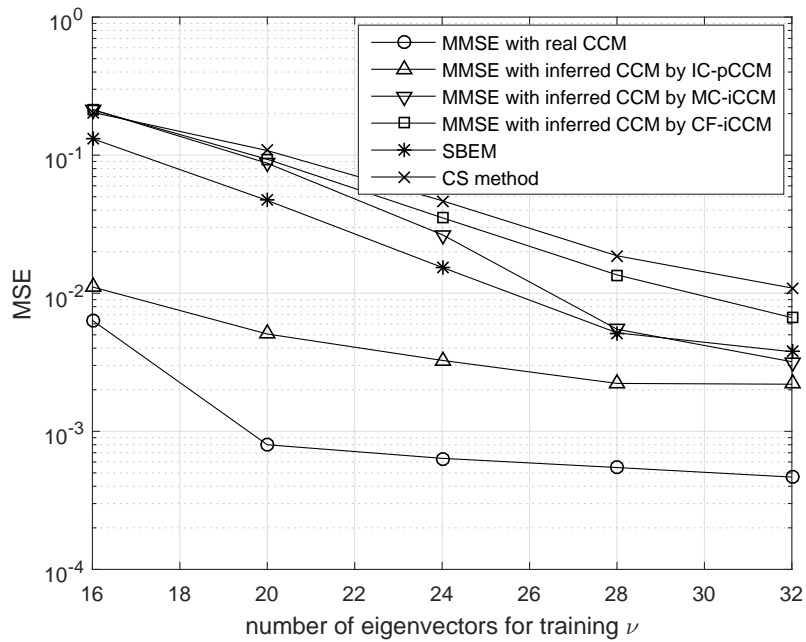}
\caption{Downlink MSE performances of MMSE estimators with reconstructed CCMs as a function of $\nu$, with AS $\Delta_{k,p}=15^\circ$, SNR=$10$ dB and uniform PAS.
\label{fig:DLMSE_tau}}
\end{figure}

As the MSE performances of MMSE estimator with CCMs inferred by different methods are significantly dependent on
the value of $\nu$, we illustrate their MSE curves as a function of $\nu$ in Fig. \ref{fig:DLMSE_tau}.
For comparison, the SBEM and CS methods with the same number of training beams are also considered.
The downlink SNR  is set as 10 dB and the AS is set as $15^\circ$.
It can be seen that the MSE curves of different methods are going to approach each other
as $\nu$ increases. This is intuitively correct  as a larger $\nu$ means a higher ratio of the channel power included for training.
It is also be seen that the MSE curves of the real CCM and the inferred CCMs by \emph{IC-pCCM} become flat quickly
when $\nu$ is about $20$, while SBEM, \emph{CF-iCCM}, \emph{MC-iCCM} and CS methods require $\nu=32$ or larger. Hence, this also indicates the
effectiveness of the \emph{IC-pCCM} scheme for accurate CCM estimation as well as accompanied low overhead channel estimation.

\section{Conclusions}
In this paper, we proposed an enhanced channel estimation scheme for TDD/FDD
massive MIMO systems based on uplink/downlink CCMs reconstruction from array signal processing techniques.
Specifically,  the CCM reconstruction was decomposed into angle estimation and PAS estimation, both of
which were extracted from only one instantaneous uplink channel estimate. With estimated angle
and PAS, uplink CCMs were first reconstructed and exploited to improve the uplink training
performances. Then the downlink CCMs were inferred  even in FDD systems.
Meanwhile, a dynamical angle division multiple access user scheduling strategy was proposed based on the real-time angle information of users.
Compared to existing methods, the proposed method does not need the long-time acquisition for uplink CCMs
and could handle a more practical channel propagation environment with larger AS.
Moreover, the proposed scheme  is applicable for any kind of PAS distributions and array geometries.
Numerical simulations have corroborated the effectiveness of our proposed scheme.


\end{document}